# BM4D-PC: nonlocal volumetric denoising of principal components of diffusion-weighted MR images

Vinicius P. Campos[1,2,*], Diego Szczupak[1], Tales Santini[3], Afonso C. Silva[1,3], Alessandro Foi[4], Marcelo A. C. Vieira[2], Corey A. Baron[5,6]

1 Department of Neurobiology, University of Pittsburgh School of Medicine, Pittsburgh, PA, USA

2 Department of Electrical and Computer Engineering, São Carlos School of Engineering, University of São Paulo, São Carlos, SP, Brazil

3 Department of Bioengineering, University of Pittsburgh Swanson School of Engineering, Pittsburgh, PA, USA.

4 Signal Processing Research Centre, Tampere University, Tampere, Finland

5 Department of Medical Biophysics, Western University, London, ON, Canada

6 Center for Functional and Metabolic Mapping, Western University, London, ON, Canada

[*] Corresponding author

E-mail: vpc24@pitt.edu

**Abstract**

**Purpose:** Noise in diffusion-weighted MRI (dMRI) is often spatially correlated due to different acquisition and reconstruction strategies, which is not fully accounted for in current denoising strategies. Thus, we propose a novel model-based denoising method for dMRI that effectively accounts for the different noise characteristics of data.

**Methods:** We propose a denoising strategy that incorporates full noise statistics, including the noise power spectral density (PSD), by leveraging the BM4D algorithm. Furthermore, to exploit redundancy across the diffusion MRI dataset, BM4D is applied to principal components (PC) of diffusion-weighted images (DWI) obtained through principal component analysis (PCA) decomposition of the entire DWI dataset, an approach we refer to as BM4D-PC. Importantly, our method also allows for direct estimation of both the noise map and PSD. We evaluated BM4D-PC against four existing state-of-the-art methods using in-silico and in vivo datasets, including high-resolution human and marmoset acquisitions.

**Results:** Overall, BM4D-PC presented the best results for the metrics PSNR, SSIM and RMSE on the in-silico experiments. The in-vivo studies also showed that BM4D-PC dramatically enhanced the image quality of raw DWIs, outperforming existing denoising methods in terms of noise suppression and detail preservation, leading to improved quality of diffusion metrics.

**Conclusion:** The proposed BM4D-PC method demonstrated state-of-the-art denoising results for dMRI using datasets from various acquisition strategies and image resolutions, potentially supporting future advances in neuroscience research.

# 1. Introduction

Diffusion-weighted MRI (dMRI) is a noninvasive imaging technique that provides unique contrast based on the diffusion of water molecules in tissue. Quantitative approaches such as diffusion tensor imaging (DTI) and diffusion kurtosis imaging (DKI) have extended the utility of dMRI for mapping white matter tracts and revealing microstructural details of the brain[1–3]. Image quality, particularly the signal-to-noise ratio (SNR), is critical for accurate data analysis. However, dMRI is inherently limited by low SNR due to the signal attenuation required for diffusion weighting. Moreover, recent hardware developments have enabled stronger diffusion weighting and high-resolution images, allowing the use of advanced mathematical methods to model tissue microstructure. Nonetheless, these methods often lead to noise propagation through non-linear computations, ultimately compromising the reliability of the derived diffusion metrics[4–6].

SNR can be improved through acquiring more diffusion directions, increasing the number of signal averages or using higher-sensitivity hardware such as advanced coils or stronger magnets[7–9]. However, these strategies lead to longer scan times and higher costs. Consequently, post-processing denoising algorithms have become attractive alternatives[10], with patch-based-PCA methods being among the most commonly used. A key example is the Marchenko–Pastur PCA (MPPCA) approach[11], which uses random matrix theory to identify and remove noise-dominated components from local patches. Several extensions have improved its flexibility and performance. NORDIC[12] denoising extends this framework by operating in the complex domain with g-factor (i.e. spatially varying noise) correction, enabling effective denoising in high-resolution and multi-band acquisitions. Tensor-MPPCA[13] adapts the method to multidimensional data (such as in multi-TE diffusion MRI) by using concepts of high-order singular value decomposition (HOSVD), and recursively denoising each dimension of the data. Ma et al. [10] incorporates the Rician variance-stabilizing transformation (VST)[14] to handle non-Gaussian noise distributions commonly found in magnitude MRI data[10]. All of these methods, while effective, typically rely on identically and independently distributed (i.i.d.) white noise[15]. However, dMRI data often undergo reconstruction steps such as zero-filling and k-space gridding, which introduce spatially correlated (colored) noise[15–18]. Henrique et al.[15] propose two variants, G-PCA and T-PCA, where they incorporate prior noise variance estimates, improving denoising performance in scenarios with colored noise. Nonetheless, their approach to estimate noise variance is limited by the use of only b=0 images and does not explicitly account for the noise power spectral density (PSD) [15].

Another class of model-based denoising methods relies on non-local block-matching strategies, which exploit the redundancy of similar image patches across space to suppress noise [19,20]. In the broader field of image denoising beyond just MRI, classical algorithms such as Non-Local Means (NLM)[19] and non-local Block-Matching and 4D transform domain filtering (BM4D)[20] have been widely adopted. In MRI, several NLM-based variants have been successfully proposed, such as the Adaptive Non-Local Means (ANLM)[21], whereas BM4D remains relatively underexplored. These block-matching methods generally underperform compared to PCA-based techniques for diffusion MRI[22] because they are typically applied independently to each 3D volume in the 4D dataset, and thus do not leverage redundancy across diffusion directions and b-values like patch-based PCA methods [22].

More recently, data-driven methods based on machine learning have emerged as powerful tools for dMRI denoising. Patch2Self[23] is a self-supervised approach that leverages the statistical independence of noise across diffusion directions to denoise without requiring clean reference data. Multidimensional Self2Self[24] extends this concept using deep convolutional networks to exploit redundancy across the multi dimensions of the data for more effective denoising. While effective, in general, machine and deep learning methods often require considerable computational resources, which can limit their applicability[25]. Furthermore, while self-supervised methods (such as the aforementioned ones) can be versatile for different datasets and acquisitions, fully supervised methods may suffer with generalizability[25]. For example, spiral dMRI is becoming increasingly feasible due to improvements in eddy current characterization[26,27], but these approaches likely have different noise and distortion characteristics (compared to EPI) which would likely preclude denoising models trained on EPI dMRI data.

In this work, we introduce a model-based denoising method for dMRI with improved incorporation of noise statistics, particularly the PSD which, to the best of our knowledge, has been ignored in dMRI denoising. We employed the BM4D[20,28] algorithm and, inspired by the approach for seismic data by Goyes-Peñafiel et al.[29], our method leverages redundancy over the 4th dimension by first computing a global PCA decomposition of a matrix comprised of vectorized diffusion-weighted images (DWIs) (each column of the matrix contains all voxels of each vectorized DWI). Then, after reshaping data, we employ BM4D[20,28] to each principal component (PC) volumetric image. This method, which we call BM4D-PC, is fully automatic, estimating noise characteristics directly from the data.

**2. Methods**

2.1 Observation model

2.1.1 White noise

For a full k-space multi-coil acquisition with Cartesian sampling, noise in the reconstructed volumetric complex-valued image can be effectively modeled as spatially varying, zero-mean, white, complex Gaussian [18]. Let $z \in \mathbb{C}$ represent this volumetric complex-valued image.

In dMRI, there is a series of $N$ DWIs $Z = [z_1, z_2, \ldots, z_N]$. Further, consider that each voxel $z_i(x) \in \mathbb{C}$ has been normalized by the spatially varying standard deviation of noise $\sigma(x)$. Note that the DWIs have a shared standard deviation $\sigma$. Then, we model the observation as:

$$z_i(x) = d_i(x) + v(x), \qquad (1)$$

where $d_i(x) = y_i(x) / \sigma(x)$ is the normalized noise-free signal and $v$ is a stationary i.i.d Gaussian white noise with unit variance [12], $v(x) = \mathcal{N}(0,1)$.

### 2.1.2 Spatially correlated (colored) noise

In practice, the k-space of the DWI is neither fully nor Cartesian sampled, such as in partial Fourier, zero-filled acquisitions, or variable-density spiral trajectories [15,18]. Therefore, noise in each DWI becomes spatially correlated[15]. We assume that noise is correlated only within each dMRI image (spatially), but not across different images. Let $q_i(x)$ now represent each DWI, the observation model becomes

$$q_i(x) = d_i(x) + \eta(x), \qquad (2)$$

where

$$\eta = v \circledast g, \qquad (3)$$

$g$ is a convolution kernel characterizing the spatial correlation of the noise, and $\circledast$ denotes convolution [28]. Since $q_i(x)$ was normalized by $\sigma(x)$, $\text{var}\{v\} = 1$, and $\text{var}\{\eta\} = \|g\|_2^2 = 1$. The correlated noise is also described by its PSD $\Psi$:

$$\Psi = E\{|\mathcal{F}[\eta_c]|^2\} = \text{var}\{\mathcal{F}[\eta_c]\} = |X| \, |\mathcal{F}[g]|^2, \tag{4}$$

where $\mathcal{F}$ is the 3-dimensional Fourier transform, and $|X|$ is the total number of voxels. Note that the DWIs have a shared PSD $\Psi$.

2.2 Patch-based PCA and singular value manipulation

The common patch-based PCA denoising approaches adopt the white noise model (Eq. 1) [11,12]. Overall, for *each 3D spatial patch*, a PCA decomposition is employed across all DWIs. Then, the denoising step consists of manipulating the obtained singular values, such that signal components will be favored in relation to noise components (e.g. thresholding). Finally, the inverse PCA of the manipulated data is performed resulting in the denoised signal[11,12].

In the thresholding strategy, only the first $k$ principal components (where $k$ is a defined threshold) are retained, presumed to correspond to signal subspace, while the remaining components, dominated by noise, are discarded. The selection of $k$ is a critical step. If $k$ is too low, important signal components may be discarded; if too high, excessive residual noise may be retained in the reconstruction. MPPCA addresses this by automatically determining $k$ based on the Marchenko–Pastur distribution of the singular values[11]. However, when the noise is spatially correlated, the assumptions underlying the Marchenko–Pastur distribution of the singular values do not hold, and its use for the automatic thresholding determination becomes unreliable[15].

2.3 BM4D-PC

Our proposal considers the correlated noise model as per section 2.1.2. Furthermore, our denoising strategy employs PCA decomposition on the entire dataset at once (global PCA). The proposed method is depicted in Figure 1 and described next.

### 2.3.1 Noise model after global PCA

Starting from (2), $q_i$ represents each 3D DWI (of size $m \times n \times o$). Then, we form a matrix of vectorized DWIs $Q = [q_1, q_2, \ldots, q_N] \in \mathbb{C}^{W \times N}$, where $W = m \cdot n \cdot o$, and $N$ is the number of diffusion-weighted image volumes. From the covariance matrix $B = Q^H Q \in \mathbb{C}^{N \times N}$, we perform eigen decomposition as

$$B = V \Lambda V^H, \tag{5}$$

where $V = [v_1, v_2, \cdots, v_N] \in \mathbb{C}^{N \times N}$ contains the orthonormal eigenvectors of $B$, and $\Lambda = \text{diag}(\lambda_1, \ldots, \lambda_N)$ contains the real-valued *eigenvalues* ordered as $\lambda_1 \geq \lambda_2 \geq \cdots \geq \lambda_N \geq 0$. The eigenvectors of $B$ are the *right singular vectors* of $Q$. Thus, right-multiplying $Q$ by $V$ yields

$$QV = A = U\Sigma, \tag{6}$$

where $U \in \mathbb{C}^{W \times N}$ contains the left singular vectors of $Q$, and $\Sigma \in \mathbb{R}^{N \times N}$ is a diagonal matrix of *singular values* of $Q$ (note that (6) also follows from the SVD of $Q = U\Sigma V^H$, but our approach is more computationally efficient).

The matrix $A$ contains the principal components of $Q$ (scaled by the singular values in $\Sigma$) and represents $Q$ expressed in the orthonormal PCA basis defined by $V$. Since each DWI contains an independent realization of the spatially correlated noise $\eta$, and because $V$ is unitary, the noise statistics are preserved through the PCA transformation. Consequently, noise is uniformly distributed across all PCs. The underlying signal, on the other hand, is concentrated in the first few components, leading to a progressively decreasing SNR across subsequent PCs.

$A$ can then be reshaped back to a 4D array of size $m \times n \times o \times N$. Each PC image $a_i$ can be written as

$$a_i = s_i + \eta, \tag{7}$$

where $s_i$ denotes the signal contribution of each PC and $\eta$ is the spatially correlated noise (Eq. 2). Consequently, the noise in $a_i$ has the same PSD $\Psi$.

2.3.2 Denoising the Principal Components

We apply BM4D[20] to each PC volumetric image $a_i$ (Eq. 7), a process we refer to as BM4D-PC. Let $\Phi(\cdot)$ represent BM4D denoising operator, then

$$\hat{s}_i = \Phi(a_i, \Psi), \tag{8}$$

where $\hat{s}_i \approx s_i$. After denoising, data is reshaped and we obtain a PC matrix $\hat{S}$, and reconstruct the denoised data matrix $\hat{D}$ as:

$$\hat{D} = \hat{S} V^H, \tag{9}$$

The columns of $\hat{D}$ are reshaped into 3D volumes, and then rescaled by the noise map $\sigma$, resulting in the final denoised DWI dataset $\hat{Y} = [\hat{y}_1, \hat{y}_2, \ldots, \hat{y}_N]$, where $\hat{y}_i \approx y_i$.

Some considerations on the denoising step:

1. **Multichannel Processing:** We employ the multichannel implementation of BM4D. Block-matching is performed only on the first PC, which has the highest SNR. The resulting block coordinates are reused across all PCs, benefiting low-SNR PCs by using high-confidence block matches.

2. **Full-Rank Denoising:** All PCs are denoised; none are discarded. Singular value manipulation could also be incorporated. However, although the smallest singular values predominantly correspond to principal components dominated by noise, both noise and signal are present in all principal components[11]. Moreover, the identification of an optimal threshold can be challenging in a global PCA context with data corrupted by spatially correlated noise.

2.3.3 Noise estimation

In section 2.3.2, we assumed prior knowledge of the noise parameters $\sigma$ (noise map) and $\Psi$ (noise PSD). An advantage of the proposed global PCA framework is that these noise parameters can be effectively estimated directly from the data. Since the PCA transformation preserves the noise statistics, the spatially varying noise pattern present in the *original*, *non-normalized* DWIs propagates identically into all PCs. Moreover, the last few PCs predominantly capture noise-only variations[30] (see Figure S1). Therefore, we use the last few PCs to compute a voxel-wise 3D noise map $\hat{\sigma}$ using a local standard-deviation estimator applied within a $5 \times 5 \times 5$ neighborhood[30].

We also estimate the noise Fourier-domain PSD $\hat{\Psi}$. First, the PCs are normalized by the estimated noise map, removing the effect of spatial variability so the resulting noise can be modeled as stationary and spatially correlated. Then, we estimate a local 2D PSD (16 x 16 voxels) by applying a local Fourier transform over a small moving window[31]. We perform it on a slice-by-slice basis, since the PSD is generally slice-invariant, primarily determined by the k-space trajectory and reconstruction algorithm. Overlapping chunks of 5 consecutive slices (step size = 3) were used, and the final local 2D PSD was obtained by taking the voxel-wise minimum across all 2D estimates. This strategy avoids overestimation, particularly in low-frequency regions, by reducing the influence of possible residual signal contamination or noise non-stationarity that may result from imperfections in the noise map estimation. Finally, the resulting local PSD is up-sampled to match the full 3D resolution of the DWI volume[31,32].

We empirically observed that using only the highest shell (b-value) leads to better noise estimation. This is likely due to the lower SNR and similar signals within the shell that reduces the signal that extends to the last several PCs. Therefore, we performed the noise estimation step using only the highest shell (Figure 1 - top). The estimation process is performed for each of the last components (three in this work), and the average is computed as a final estimate.

2.3.4 BM4D

BM4D exploits non-local self-similarity within a volumetric image[20]. Overall, it follows a two-stage procedure. In the first stage, a basic estimate of the noise-free image is generated. For each reference voxel, a block is defined, and similar blocks within a search region are grouped into a 4D array. A separable 4D wavelet transform is then applied, yielding a sparse representation of the signal. Noise is suppressed by *hard thresholding* the wavelet coefficients, with a threshold proportional to the transform-domain noise variances of the grouped blocks [20,28]. Finally, an inverse 4D wavelet transform is applied to reconstruct the denoised blocks.

In the second stage, the "pre-denoised" image from the first stage is used to improve both block matching and filtering. Moreover, at this stage, Wiener filtering replaces hard thresholding: the grouped blocks transform coefficients are weighed according to their noise variances and the energy of the corresponding pre-denoised transform coefficients. After inverse wavelet and aggregation of the denoised blocks, the result is a more accurate final denoised image.

BM4D was originally developed for Gaussian i.i.d. white noise[20], but it has been extended to handle stationary spatially correlated noise[28]. It incorporates the noise PSD into the process, and calculates exact transform-domain noise variances of the grouped blocks, which then significantly improves block-matching, shrinkage accuracy on transform domain, and aggregation of the denoised blocks, resulting in effective denoising of spatially correlated noise[28].

2.4 In silico validation

We used a publicly available noise-free diffusion MRI simulated dataset to validate the approach with a known ground-truth[10] (https://github.com/XiaodongMa-MRI/Denoising). The data was simulated with Fiberfox[33], based on the ISMRM 2015 Tractography Challenge brain phantom[34]. The dataset includes 67 images (7 b = 0, 30 b=1000 and 30 b=2000 s/mm2) simulated at 2mm isotropic resolution. To generate complex-valued data, we first added smooth phase variations to the noise-free magnitude images. Additionally, random global phase shifts were applied independently to each DWI volume.

Complex Gaussian noise with spatial variation was added at three reference levels: 1%, 5%, and 10% of the maximum signal intensity of the b=0 image. We simulated both white and colored noise; for the latter, spatial correlation was introduced by convolving the noise image with a band-pass filter kernel, resulting in a non-flat PSD (Figure S1).

Denoising performance was quantitatively assessed using Peak Signal-to-Noise Ratio (PSNR) and Structural Similarity Index Measure (SSIM)[35]. PSNR is defined as:

$$\text{PSNR} = 10 \log_{10}\left(\frac{\max(y)^2}{\frac{1}{N}\sum_{i=1}^{N}(y_i - \hat{y}_i)^2}\right), \qquad (10)$$

where $y$ and $\hat{y}$ represent the ground-truth and denoised images respectively, and $N$ is the total number of voxels.

After denoising, we fitted a DTI model to the data using only the b=1000 shell using the DIPY library[36]. DKI was also performed using all shells, following an axisymmetric fitting model[37,38] without spatial regularization, implemented in the MatMRI toolbox[39]. Spatial regularization was intentionally omitted to isolate the impact of denoising from smoothing effects during model fitting. We then extracted diffusion metrics fractional anisotropy (FA), mean diffusivity (MD), and mean kurtosis (MK) and computed the root mean squared error (RMSE) for each metric relative to the noise-free reference.

2.5 In vivo validation - human

2.5.1 EDDEN[40] dataset

We used a publicly available diffusion MRI dataset[22,40], acquired at 0.9 mm isotropic resolution (TR=6.569 s, TE=91 ms, multiband factor = 3, in-plane GRAPPA = 2). Only one repeat was used, and it contains 202 volumes (14 b=0, 93 b=1000, and 92 b=2000 s/mm²) with AP phase encoding direction; and 3 *b*=0 with PA phase encoding. This dataset represents an EPI acquisition with very low SNR due to its ultra-high spatial resolution.

Following denoising, diffusion MRI preprocessing was performed using Mrtrix3[41] and FSL[42], including topup, motion correction, and eddy current-induced geometric distortions correction. DTI and DKI were fit as described in section 2.4.

2.5.2 In house dataset (µFA)

A healthy volunteer was scanned at Western University's Center for Functional and Metabolic Mapping (CFMM). This study was approved by the institutional review board at Western University, and informed consent was obtained before scanning. Diffusion MRI was acquired using a single-shot variable-density spiral trajectory on a 3T Siemens PRISMA Fit scanner with field probe monitoring[43–45]. Both linear tensor encoding (LTE) and spherical tensor encoding (STE) were used[46,47], with acquisition parameters: TE/TR=80/9000 ms, multiband factor = 3, in-plane acceleration = 3, and 1.2 mm isotropic resolution. The diffusion protocol included b-values of 0 (6 LTE), 150 (6 LTE), 1000 (26 LTE), and 2000 s/mm² (30 STE + 26 LTE). This high-resolution, non-Cartesian dataset is well-suited for highlighting the strong spatial correlation of noise, serving as a highly valuable test case for assessing denoising performance.

For this dataset, no distortion correction was required, as a field-probe-informed reconstruction was performed using MatMRI[39]. Following denoising, DKI was fitted using a b-tensor free water elimination model[48]. It is important to note that this model yields kurtosis metrics that are highly sensitive to noise. From this analysis, total, spherical, and anisotropic kurtosis maps ($K_{total}$, $K_{iso}$ and $K_{aniso}$ respectively) were derived.

2.6 In house - In vivo - marmoset

For the last experiment, in vivo marmoset data were collected at University of Pittsburgh. All animal procedures in this study were approved by the Animal Care and Use Committee of the University of Pittsburgh (IACUC protocol #24014391). Diffusion MRI was acquired on a 9.4T Bruker scanner using a 2D spin-echo EPI sequence with parameters: TR=5.1s, TE=38ms, in-plane GRAPPA = 2, partial Fourier = 1.25, and 0.5mm isotropic resolution. The images were acquired for 2 phase encoding directions, and each has 3 shells (16 b=0, 64 b=1000, and 128 b=2000 s/mm2).

The coil-specific images obtained from the vendor GRAPPA reconstruction were combined in MATLAB using the SENSE1[49] method to obtain complex images, where the sensitivity profiles of the coils were estimated using ESPIRIT[50] within BART toolbox[51,52]. After denoising, diffusion MRI preprocessing, DTI fit, and DKI fit were performed as described in section 2.5.1. The main purpose of this dataset is to evaluate the denoising in terms of generalization to preclinical subjects/scanners.

2.7 – Implementation and benchmarking

We implemented BM4D-PC in MATLAB and the code is available at https://github.com/viniciuspcampos/BM4D-PC. Our method was compared against four others: MPPCA[11], using the implementation of Olesen, J.L. et al.[13], available in DESIGNER-v2 (https://nyu-diffusionmri.github.io/DESIGNER-v2/)[53,54]; NORDIC[12] (https://github.com/SteenMoeller/NORDIC_Raw); Patch2Self[23], available in DIPY[36] (https://dipy.org/); and Threshold PCA (T-PCA)[15], based on the authors' implementation (https://github.com/RafaelNH/PCAdenoising). Main parameters for all methods are described in Table S3.

To further evaluate our proposed strategy, for the in-silico dataset, we compared BM4D-PC with the standard BM4D, applied independently to each DWI volume.

2.8 - Complex-valued data

All experiments used complex-valued data. Before denoising, we applied slice-by-slice phase stabilization[12,22,55]. Briefly, the phase of each slice in each volume was estimated using a low-pass filter and was then removed by approximately rotating the complex signal toward the real axis. The imaginary component, primarily containing noise, was discarded, and only the real part was retained for denoising. This preprocessing step was applied for all denoising methods.

**3. Results**

3.1 – In silico

Colored noise

Table 1 summarizes the results of DWI image quality metrics. While all methods improved image quality, BM4D-PC consistently achieved the best results for both PSNR and SSIM, except for b=2000s/mm$^2$ at the 1% noise level. Notably, the greatest improvements were observed at the highest noise level (10%), where the gain in PSNR for BM4D-PC was 58.11%, 155.68% and 142.91% for b=0, b=1000 and b=2000s/mm$^2$ respectively. The next closest performance for each b-shell was 39.78% (NORDIC), 111.01% (MPPCA), and 125.39% (MPPCA) for the same three b-shells.

To qualitatively assess the denoising performance, an axial slice of a raw DWI image at 5% noise level is presented (Figure 2a). All methods visibly improve upon the noisy input; however, BM4D-PC outperformed the others. For the b=2000s/mm$^2$ images, the visual improvement is especially pronounced. The voxel-wise absolute error maps (Figure S2) further highlight the efficacy of the methods, and overall BM4D-PC presented fewer visual errors (darker maps). All the methods show little structure in the noise residuals (Figure S2).

The improvement in DWI quality translates into enhanced diffusion metrics. Our method achieves the lowest RMSE for the very low SNR experiments (Table 2 – 5% and 10% noise-levels), while presenting higher RMSE than others at 1% noise-level. Visually, all methods achieve better diffusion metrics compared to the noisy input (Figure 2b), with BM4D-PC presenting the best results.

Finally, BM4D-PC also outperforms the standard BM4D approach for both raw DWI and diffusion metrics (Figure S3), which stems from exploiting redundancy along the fourth dimension of the dataset via global PCA.

White noise

Table S1 summarizes the DWI image quality metrics for the white noise scenario. Similar to the colored noise case, BM4D-PC achieved the best results overall. Visual results are summarized in Figure S4, in which we show DWIs related to b-values=1000 and 2000s/mm$^2$ (Figure S4a) and the diffusion metrics (Figure S4b). Again, the improvements in DWI are translated to improved diffusion metrics. Finally, Table S2 summarizes the RMSE with respect to diffusion metrics, where BM4D-PC presents the lowest RMSE at the highest noise levels (5 and 10%).

3.2 – In vivo Human

3.2.1 – EDDEN dataset

Similar to the in-silico case, all the denoising methods improved the quality of the DWIs for the in vivo human EDDEN dataset (Figure 3 - top). In terms of residuals (difference between the noisy and the denoised images), no difference can be seen among the methods. Visually, BM4D-PC outstands by effectively suppressing noise while preserving anatomical details. As for the diffusion metrics (Figure 3 - bottom), the visual discrepancy in favor of BM4D-PC is even more pronounced, with special attention towards the parallel kurtosis ($K_{||}$).

A numerical evaluation was also performed, where mean and standard deviation of the FA values were calculated in two regions of interest: ventricles, and midbody of corpus callosum (CC) (Figure 4), where FA is expected to present very low and high values, respectively. For the ventricles (Figure 4a), BM4D-PC achieves the lowest mean value, while also presenting the lowest standard deviation. For the CC region (Figure 4b), BM4D-PC, MPPCA and T-PCA achieve similar high mean values, with MPPCA being slightly higher. However, BM4D-PC presents the smallest standard deviation.

3.2.2 – In house dataset (µFA)

A similar visual analysis of the µFA spiral dataset (Figure 5) shows notable DWI quality improvement for all methods. BM4D-PC outperforms all the others, recovering small details and drastically suppressing noise. The kurtosis metrics (Figure 5 – bottom) further highlight the superiority of our method, especially in isotropic kurtosis (Kiso), which is highly sensitive to noise.

3.3 In house - In vivo - marmoset

Finally, the results of the marmoset data are presented (Figure 6). Note how strong the noise spatial correlation is, with a clear vertical pattern (Figure 6 – top). The improvement in DWI quality is achieved by Patch2Self, T-PCA and BM4D-PC, while MPPCA and NORDIC did not perform well, almost not suppressing any noise. These results are translated to the diffusion metrics (Figure 6 – bottom), where BM4D-PC presents visually slightly better results when compared to both T-PCA and Patch2Self.

**4. Discussion**

In this work, we proposed BM4D-PC, a novel model-based denoising method for diffusion MRI data. Our method can characterize noise directly from the data and use it during denoising, introducing the explicit estimation and use of the noise PSD for the first time for dMRI. Notably, the noise PSD reflects k-space sampling characteristics, such as the oversampling near the center of k-space for variable density spiral (Figures 7 and S6). Additionally, we proposed denoising the Principal Component images obtained from a global PCA of the entire DWI dataset, rather than the DWIs. We conducted simulation studies and were able to demonstrate the efficacy of BM4D-PC for both colored and white noise. When benchmarked with other state-of-the-art dMRI denoising methods (Patch2Self[23], MPPCA[11,13], NORDIC[12], and T-PCA[15]), our proposal exhibited, overall, the best results (Figure 2; Figures S2-S5; Tables 1- 2;Tables S1-S2). We also performed several in-vivo human and marmoset studies. In accordance with the simulation, BM4D-PC dramatically enhanced the image quality of raw DWIs, leading to outstanding diffusion metrics results, outperforming the existing dMRI denoising methods (Figure 3-6).

4.1 Differences from Goyes-Peñafiel et al.[29]

Our work was inspired by Goyes-Peñafiel et al.[29]. We highlight, though, key differences to their method[29]: (1) the application to MRI is greatly different than seismic data; (2) the authors[29] use the denoiser as a regularizer on solving the problem of recovering missing seismic data, while here we propose BM4D-PC for image denoising; (3) due to their application, there is no noise estimation proposal, neither of the noise map nor of the noise PSD; (4) the noise is assumed to be stationary and white throughout their regularization process, while the ability of BM4D to account for colored noise plays a crucial role in our denoising proposal.

4.2 Multichannel approach

We adopted the multichannel BM4D strategy, where block-matching is performed only once, on the first PC, and the positions are reused on the remaining PCs. The advantages of this approach are two-fold. First, the similarity calculated between blocks is more reliable on the first PC, due to its high SNR. Therefore, the low-SNR PCs take advantage of the high-quality block-matching estimation, leading to improved denoising results. Second, the block-matching step can be time-consuming, and by doing only once we speed up the overall denoising procedure.

The main trade-off is that blocks matched on the first PC may, in *rare* cases, correspond to *highly* dissimilar blocks in other PCs. In this scenario, the resulting loss of sparsity in the transform domain can reduce noise suppression efficiency; nevertheless, this limitation does not completely undermine BM4D, as structural differences are preserved. Noise suppression is applied after shrinkage of the transform-domain coefficients of grouped blocks, which ensures that large coefficients, associated with true structural differences are retained, ensuring that differences between blocks are preserved after inverse transform[56].

4.3 Generalizability

Like other methods, the BM4D-PC algorithm includes configurable parameters, mainly those related to the denoising algorithm itself, BM4D. All the studies performed here used the same parameters (Table S3), proving how versatile BM4D-PC is in very different scenarios. Regarding the results for the in-silico data at 1% noise level (Tables 1-2, S1-S2), our method showed slightly higher RMSE for some diffusion metrics (MD and MK). We point out, though, that such low noise level is not typically encountered in actual dMRI scanning (especially at high resolution), and when it is, denoising methods will generally offer only limited benefit. Furthermore, while parameter tuning would have improved results, we kept them fixed to allow consistent comparisons across noise scenarios.

Similarly, the model-based PCA methods needed little or no tuning (Table S3). However, the opposite was observed with respect to Patch2Self, where the choice of the regressor can be cumbersome (see Fadnavis et al.[23] for further details). We tried several options and none of the linear regressors would provide satisfactory results. Thus, we used the multilayer perceptron regressor, based on the implementation of Kang et al.[24] (Table S3). The in-silico results show the good performance of this setup; nevertheless, it was not satisfactory for the in-vivo EDDEN dataset (Figures 3-4), suggesting it would need more tuning and troubleshooting, which can be challenging and is outside the scope of this work.

It is important to note the different performances of MPPCA, NORDIC and T-PCA across datasets. T-PCA demonstrated limited performance on the in-vivo µFA dataset (Figure 5), and underperformed MPPCA and NORDIC on the in-silico data (Figures 2 and S4). This may be attributed to its noise estimation strategy, which relies on the b=0 images. Only 7 b=0 were available in the in-silico dataset and 6 in the µFA dataset, compared to 14 in EDDEN and 16 in the marmoset datasets. A more accurate noise estimation strategy, potentially the one presented here, could improve T-PCA's performance, as already suggested by the authors[15]. Finally, MPPCA and NORDIC failed to suppress noise in the marmoset data (Figure 6), probably misclassifying all singular values as related to signal components. This agrees with the findings of the T-PCA study[15], confirming that under strong noise correlation those methods struggle to perform effective denoising.

### 4.4 Magnitude vs complex-valued data

We focused on denoising complex-valued MRI data, where noise is zero-mean and Gaussian distributed, as opposed to magnitude images, which exhibit non-zero-mean Rician noise[14,18]. Nevertheless, our method may be also suitable for magnitude-only data under the use of the Rician VST[10,14], that transforms Rician distributed data into approximately Gaussian distributed. This approach for denoising dMRI data has been well demonstrated by Ma et al.[10].

### 4.5 Drawbacks and improvements

One drawback of BM4D-PC is the processing time that can scale up fast with data size, where the main burden can be attributed to the denoising step. For example, on the in-vivo EDDEN dataset, BM4D-PC took approximately 1 hour, compared to less than 4 minutes for MPPCA. However, it is still considerably faster than Patch2Self, which took nearly 9 hours. Similarly, the memory consumption required for BM4D-PC can be high, also due to global PCA computation. Therefore, future code enhancements can take place, specially targeting the use of GPU (not used in our work). Furthermore, even though our method provides reliable and accurate estimates of the noise map and PSD, improving the estimation step may result in better denoising performance.

### 4.6 Future avenues

While we demonstrated our method in the context of diffusion MRI, BM4D-PC is likely suitable for other high-dimensional MRI modalities, such as functional MRI, and dMRI with multiple TEs, TR or inversion times. Nevertheless, detailed investigation needs to be carried out and may be a matter of future work. Ultimately, we believe BM4D-PC will enhance data quality in a broad range of applications, especially the ones that incorporate high-resolution data and advanced diffusion metrics, allowing for future advances in neuroscience research.

## 5. Conclusion

In this work, we presented BM4D-PC, a novel model-based method for denoising dMRI images. BM4D-PC outperformed state-of-the-art methods for diffusion MRI applications with various acquisition strategies and resolutions, including high-resolution data with strong spatially correlated noise.

**Data availability statement**

The source code of BM4D-PC is publicly available at https://github.com/viniciuspcampos/BM4D-PC . The acquired in vivo data are also available upon request.

**Acknowledgments**

This study was financed in part by the Coordenação de Aperfeiçoamento de Pessoal de Nível Superior - Brasil (CAPES) - Finance Code 001, and by the São Paulo Research Foundation (FAPESP), grants number 2021/12673-6 and 2025/00190-1. It was also financed by grants R24AG073190 and U19AG074866 from the National Institutes of Health, National Institute on Aging. It was also funded by Research Council of Finland, PROFI6 Tampere Imaging Platform (project no. 336357). This work was also supported by the Natural Sciences and Engineering Research Council of Canada, grant RGPIN-2018-05448.

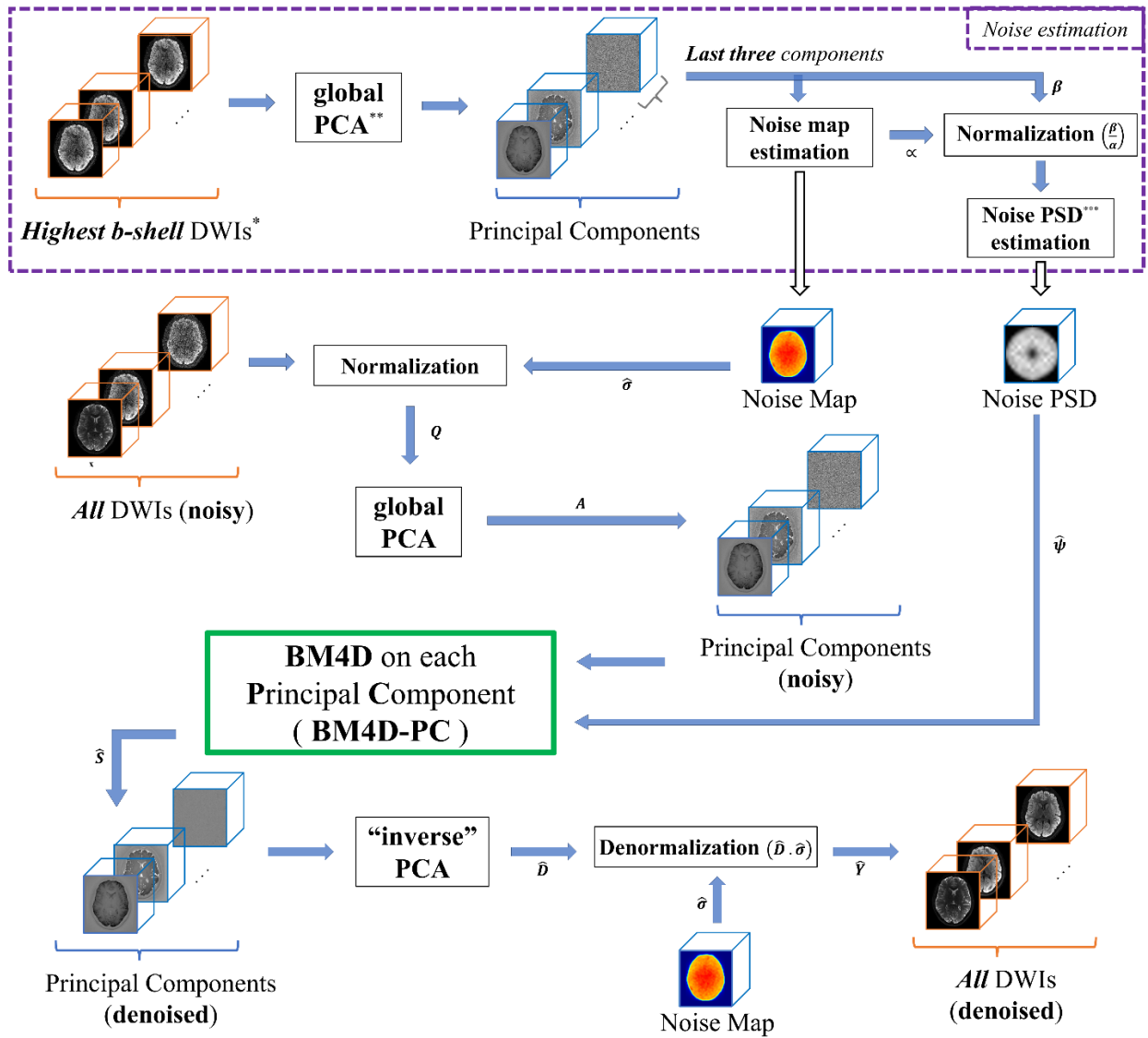

*Figure 1* - *Framework of the proposed denoising method. The top section (dashed rectangle) represents the noise estimation part, where Noise Map and Noise PSD are obtained and then used as inputs to the subsequent steps of the denoising framework.*

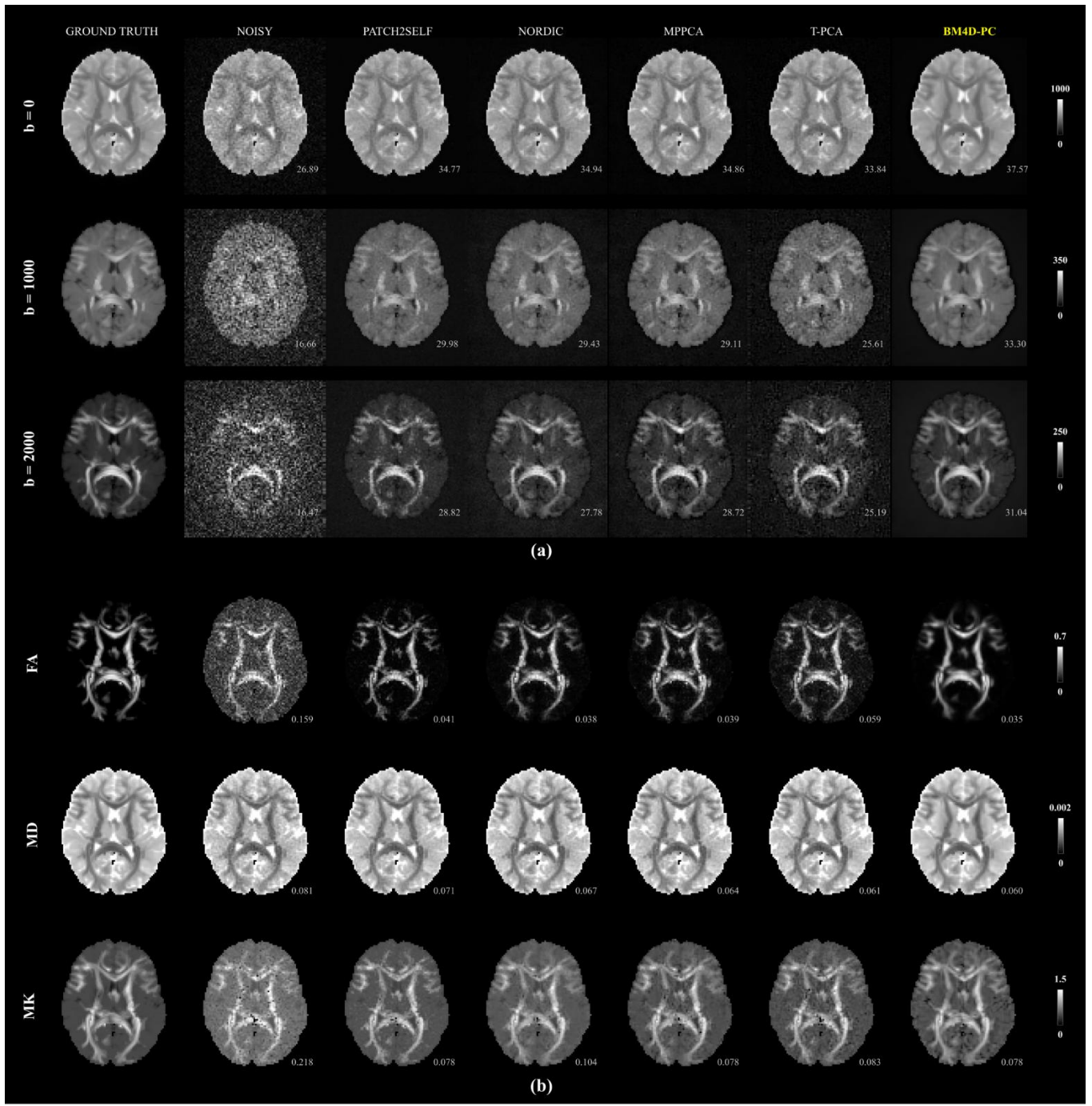

*Figure 2* – In silico experiment, colored noise at 5% noise level. (a) Results of a representative slice of a single DWI, in which we show Ground Truth and Noisy images, followed by the denoised version of each method. The numbers on each image represent the PSNR (dB) of the 3D volume. (b) DTI metrics FA, MD, and DKI metric MK. The numbers on each image represent the RMSE with respect to the Ground Truth.

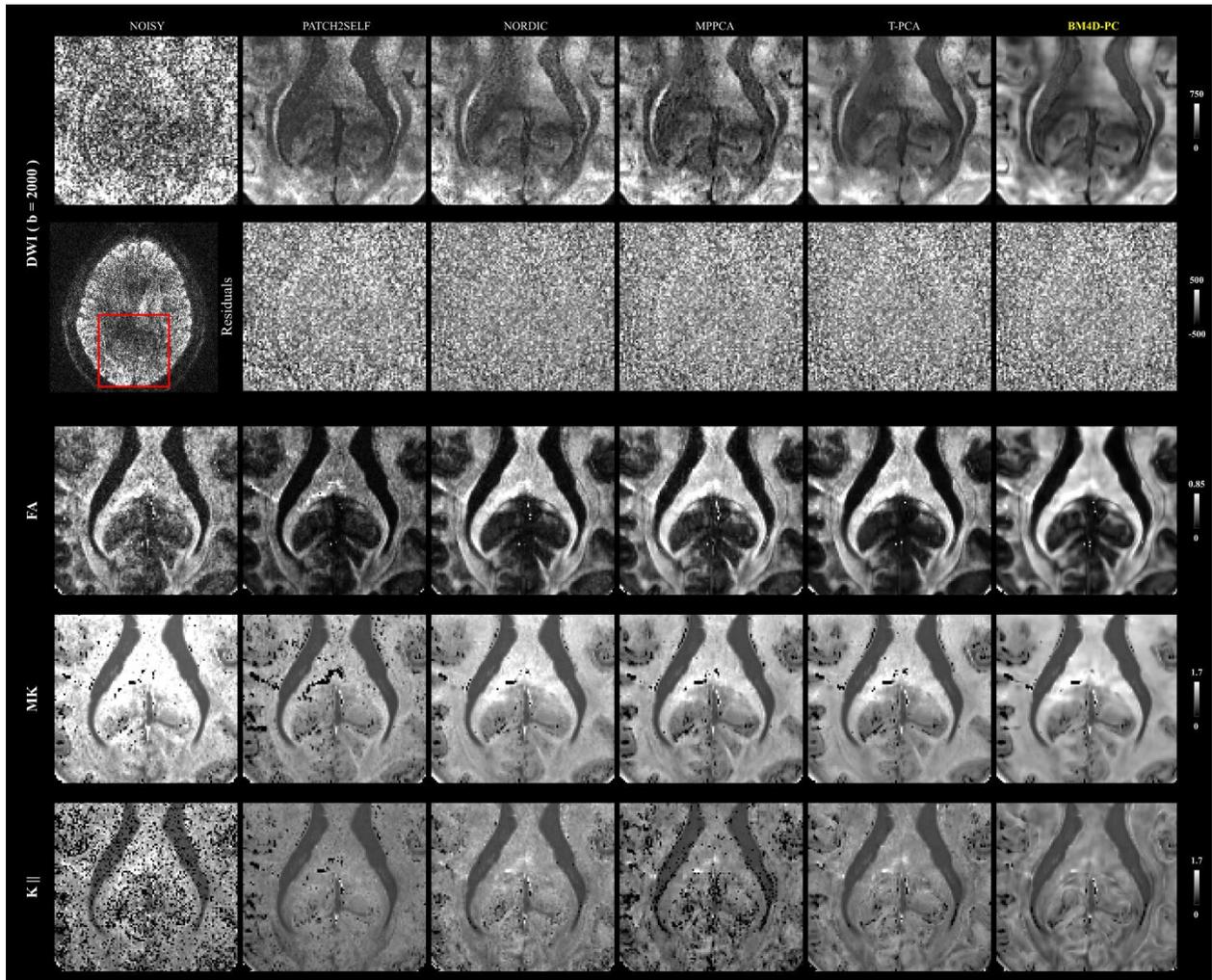

***Figure 3*** *– In vivo experiment of human dataset EDDEN. (Top section) We show a zoomed region (highlighted by the red rectangle) of a representative slice of a single raw DWI (b-value = 2000) along with the residuals (difference between Noisy and the corresponding denoised result of each method). (Bottom section) The corresponding diffusion metrics FA, MK and K|| (parallel kurtosis).*

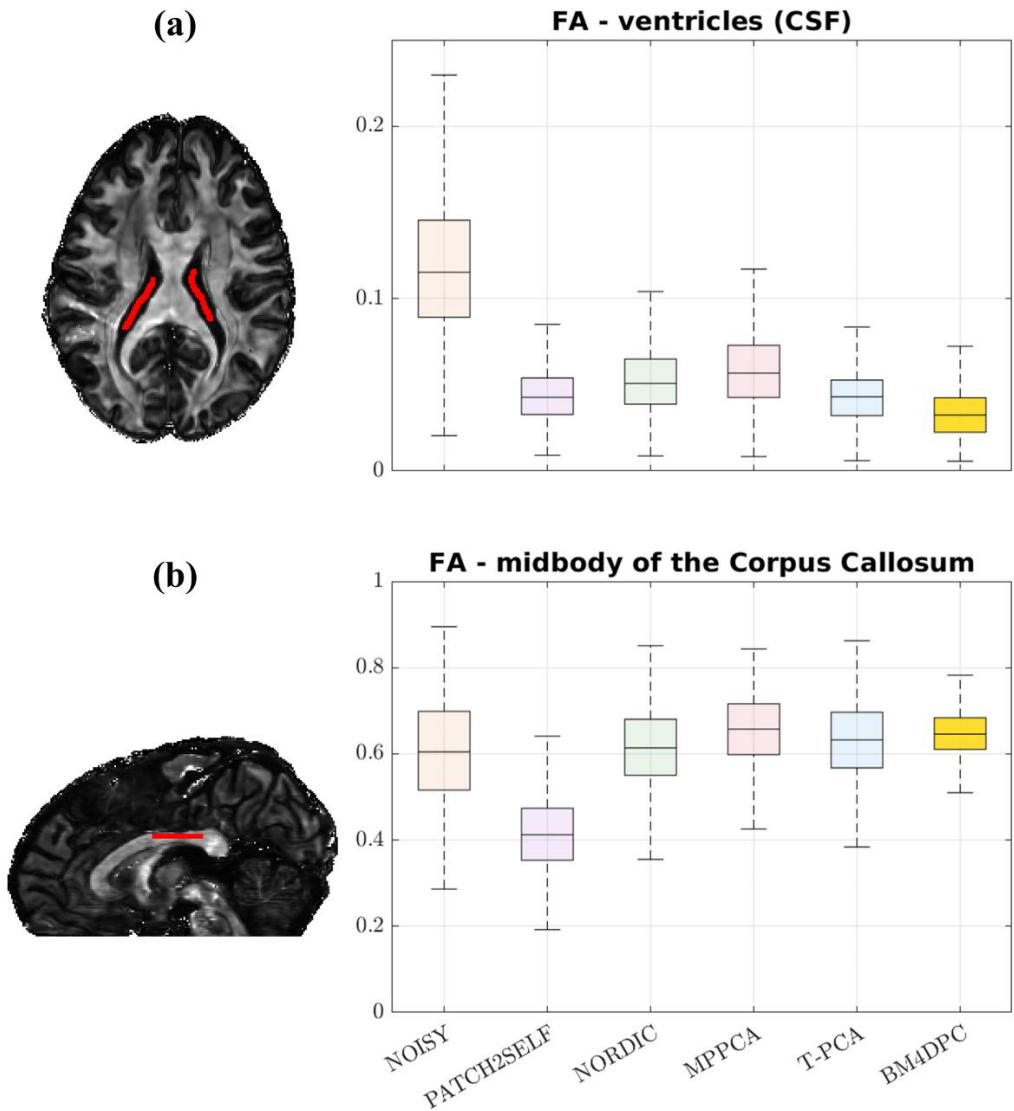

*Figure 4* – Numerical assessment of in vivo human results of EDDEN dataset. (a) Boxplot of FA values calculated in delimited ROI highlighted in red, which corresponds to the ventricles region. (b) Boxplot of FA values calculated in the delimited ROI highlighted in red, which corresponds to the midbody of Corpus Callosum (CC) region. Each boxplot corresponds to a different method, as per labels on the x-axis. BM4D-PC achieves the lowest mean FA in the CSF region, and high values on the CC region, in accordance with expected. In both cases, it presents the lowest standard deviation.

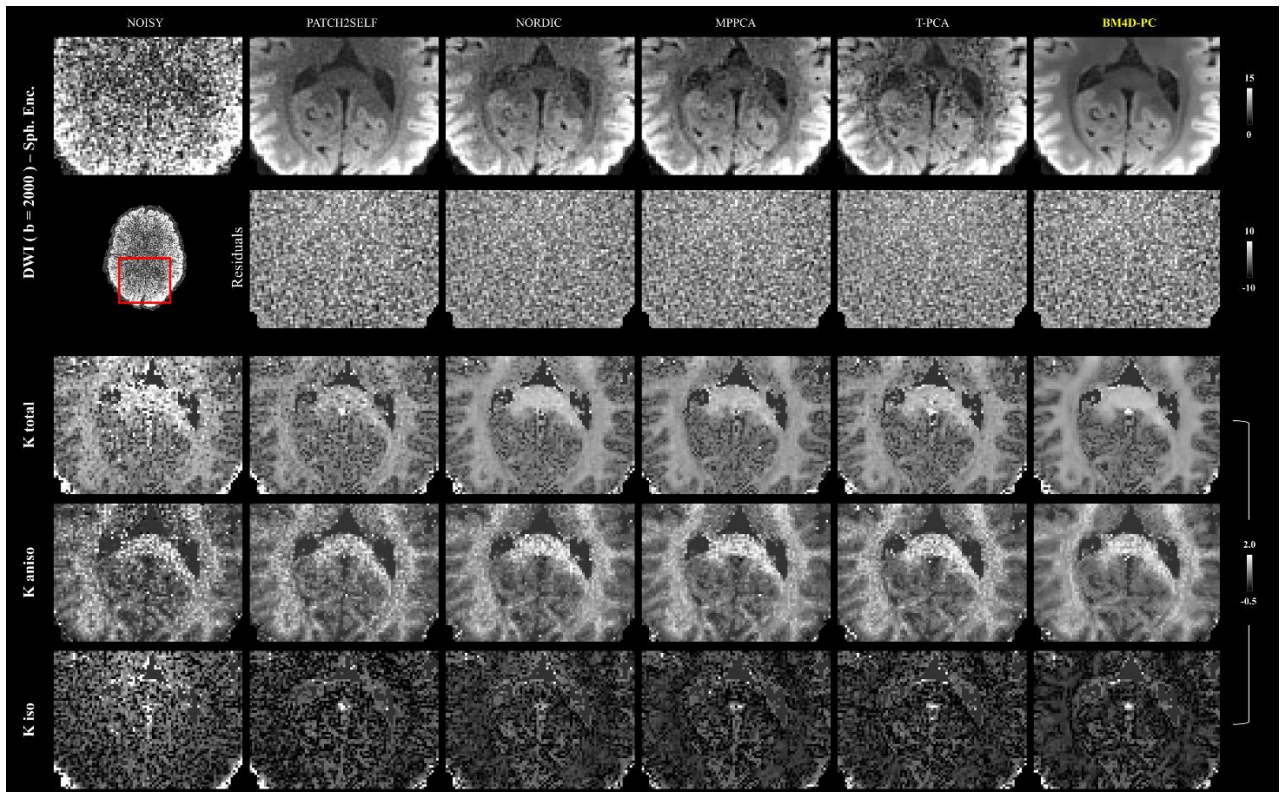

*Figure 5* - In vivo experiment of human dataset acquired with spiral trajectory. (Top section) We show a zoomed region (highlighted by the red rectangle) of a representative slice of a single raw DWI, acquired with spherical encoding and b-value = 2000. We also present the residuals (difference between Noisy and the corresponding denoised result of each method). Note how BM4D-PC achieves superior noise suppression while preserving details. (Bottom section) The corresponding diffusion metrics Ktotal (total kurtosis), Kaniso (anisotropic kurtosis), and Kiso (isotropic kurtosis).

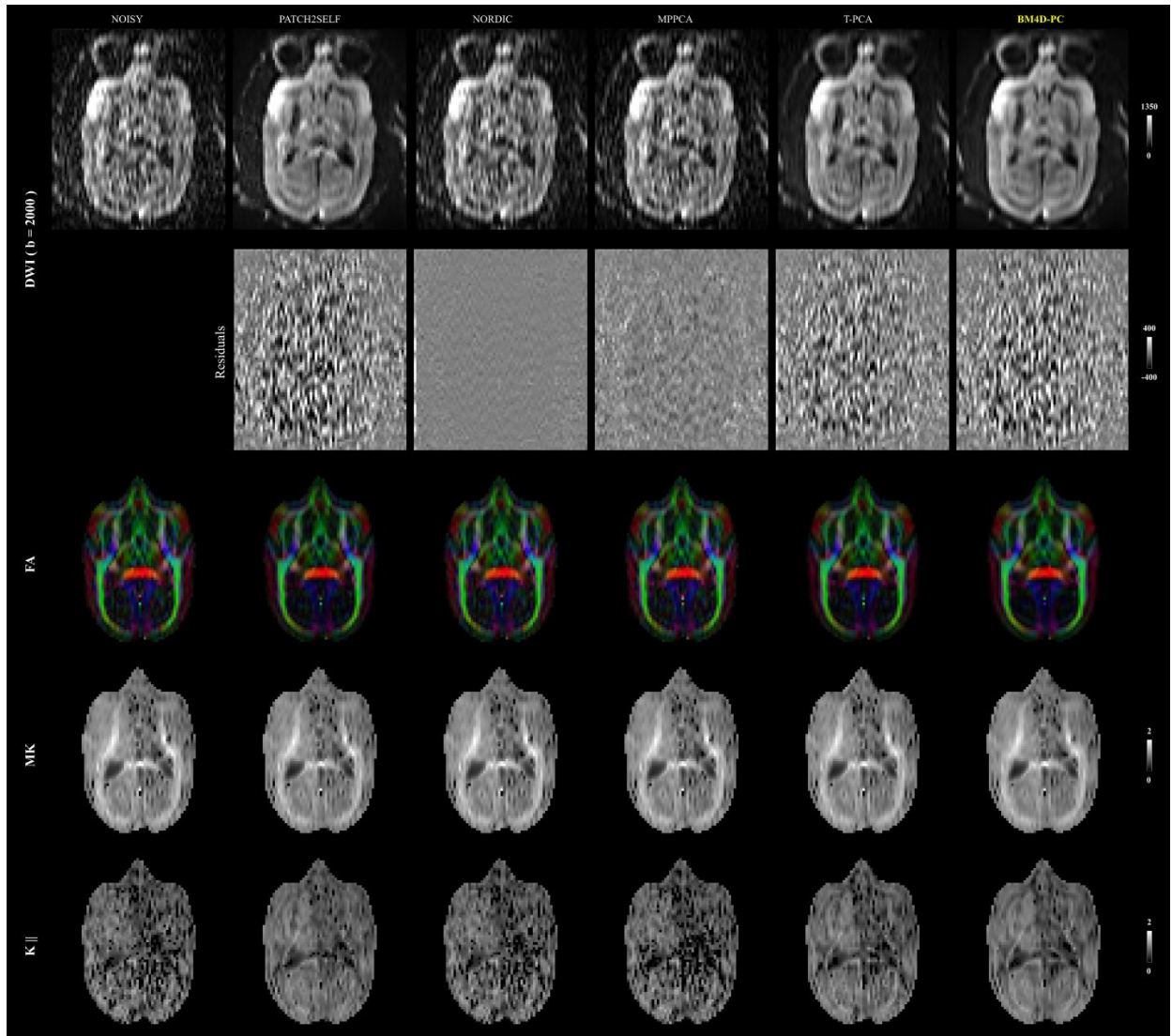

***Figure 6*** - *In vivo experiment of marmoset dataset. (Top section) We show a representative slice of a single raw DWI (b-value = 2000) along with the residuals (difference between Noisy and the corresponding denoised result of each method). (Bottom section) The corresponding diffusion metrics FA, MK and K|| (parallel kurtosis).*

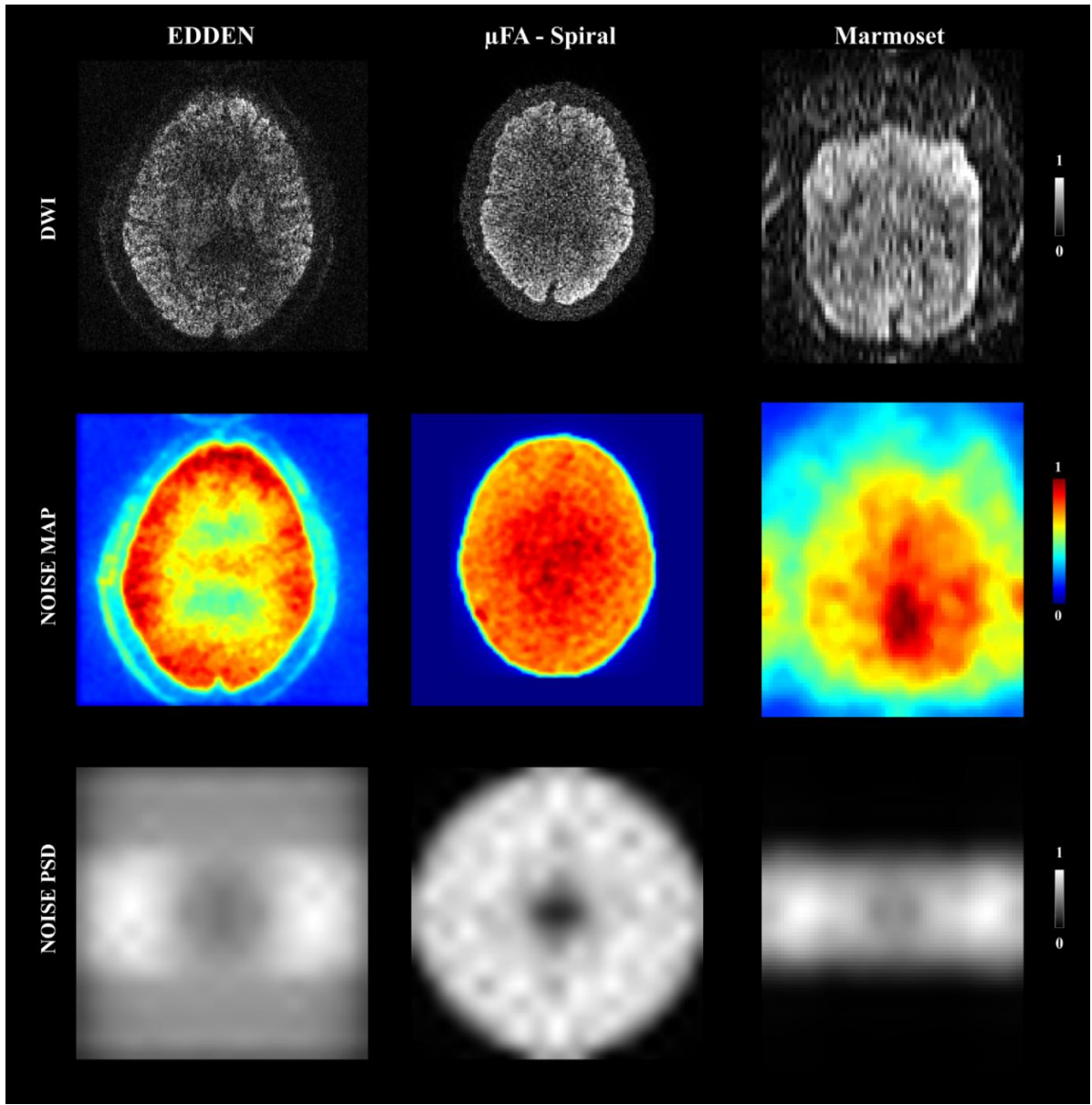

*Figure 7* –Noise Maps and Noise PSDs obtained by our proposed estimation framework for the three in-vivo datasets used in this work. (Top) A noisy raw DWI is shown for reference. The shadowed black region in the center of the PSDs, more evident on EDDEN dataset (left column - bottom), is related to the phase stabilization step, in which a low pass filter is used. The estimated noise maps and PSDs relate to the acquisition and reconstruction strategy and to the noise pattern, with a typical g-factor noise map pattern for EDDEN (related to in-plane GRAPPA acceleration), a PSD mirroring the variable density spiral used for the middle column, and a PSD mirroring the zero filling used for the marmoset data. Importantly, for the marmoset data, zero-filling was done on the frequency encoding dimension (vertical), whereas partial Fourier was performed on the phase encoding dimension (horizontal). Notably, only the real part of the phase-aligned data is being used, and, therefore, even for the partial Fourier acquisition, the PSD is conjugate-symmetric. When complex data is used for the PSD estimation, a non-symmetric partial Fourier sampling pattern is evident (see Figure S6).

|  |  | NOISE LEVEL 1% | | | NOISE LEVEL 5% | | | NOISE LEVEL 10% | | |
|---|---|---|---|---|---|---|---|---|---|---|
|  |  | b = 0 | b = 1k | b = 2k | b = 0 | b = 1k | b = 2k | b = 0 | b = 1k | b = 2k |
| **PSNR** | NOISY | 40.86 | 31.03 | 30.79 | 26.87 | 17.22 | 16.99 | 20.89 | 11.53 | 10.16 |
|  | PATCH2SELF | 48.09 | 41.20 | 40.19 | 34.77 | 30.38 | 29.02 | 29.17 | 24.08 | 21.20 |
|  | NORDIC | 48.56 | 41.67 | **40.86** | 34.95 | 29.83 | 28.22 | 29.20 | 24.20 | 20.31 |
|  | MPPCA | 48.46 | 41.49 | 40.61 | 34.86 | 29.53 | 29.19 | 28.99 | 24.33 | 22.90 |
|  | T-PCA | 47.40 | 39.40 | 38.42 | 33.85 | 26.12 | 25.70 | 28.08 | 20.44 | 19.86 |
|  | BM4D-PC | **49.46** | **42.75** | 40.46 | **37.57** | **33.52** | **31.34** | **33.03** | **29.48** | **24.68** |
| **SSIM** | NOISY | 0.984 | 0.852 | 0.809 | 0.753 | 0.323 | 0.257 | 0.511 | 0.158 | 0.085 |
|  | PATCH2SELF | 0.997 | 0.983 | **0.983** | 0.940 | 0.858 | 0.831 | 0.829 | 0.625 | 0.458 |
|  | NORDIC | 0.997 | 0.985 | 0.982 | 0.943 | 0.830 | 0.827 | 0.826 | 0.632 | 0.576 |
|  | MPPCA | 0.997 | 0.985 | 0.981 | 0.942 | 0.825 | 0.822 | 0.823 | 0.625 | 0.600 |
|  | T-PCA | 0.997 | 0.975 | 0.962 | 0.931 | 0.704 | 0.641 | 0.802 | 0.458 | 0.377 |
|  | BM4D-PC | **0.998** | **0.989** | 0.981 | **0.973** | **0.930** | **0.902** | **0.934** | **0.880** | **0.793** |

*Table 1* – *In silico experiment for colored noise. Quantitative metrics of the DWIs are separated by noise level and b-value. Best results in bold.*

|  |  | NOISE LEVEL 1% | NOISE LEVEL 5% | NOISE LEVEL 10% |
|---|---|---|---|---|
| **FA** | NOISY | 0.031 | 0.159 | 0.266 |
|  | PATCH2SELF | 0.012 | 0.041 | 0.082 |
|  | NORDIC | **0.010** | 0.038 | 0.063 |
|  | MPPCA | 0.011 | 0.039 | 0.065 |
|  | T-PCA | 0.013 | 0.059 | 0.111 |
|  | BM4D-PC | 0.011 | **0.035** | **0.052** |
| **MD** ($\times 10^{-3}$) | NOISY | 0.015 | 0.081 | 0.199 |
|  | PATCH2SELF | 0.014 | 0.071 | 0.136 |
|  | NORDIC | 0.014 | 0.067 | 0.143 |
|  | MPPCA | 0.014 | 0.064 | 0.126 |
|  | T-PCA | **0.013** | 0.061 | 0.120 |
|  | BM4D-PC | 0.016 | **0.060** | **0.105** |
| **MK** | NOISY | 0.038 | 0.218 | 0.457 |
|  | PATCH2SELF | 0.024 | **0.078** | 0.171 |
|  | NORDIC | 0.023 | 0.104 | 0.233 |
|  | MPPCA | 0.022 | **0.078** | 0.164 |
|  | T-PCA | **0.021** | 0.083 | 0.169 |
|  | BM4D-PC | 0.032 | **0.078** | **0.158** |

*Table 2* – *In silico experiment for colored noise. Root Mean Squared Error (RMSE) values for the diffusion metrics. Best results in bold.*

# Supplementary Material

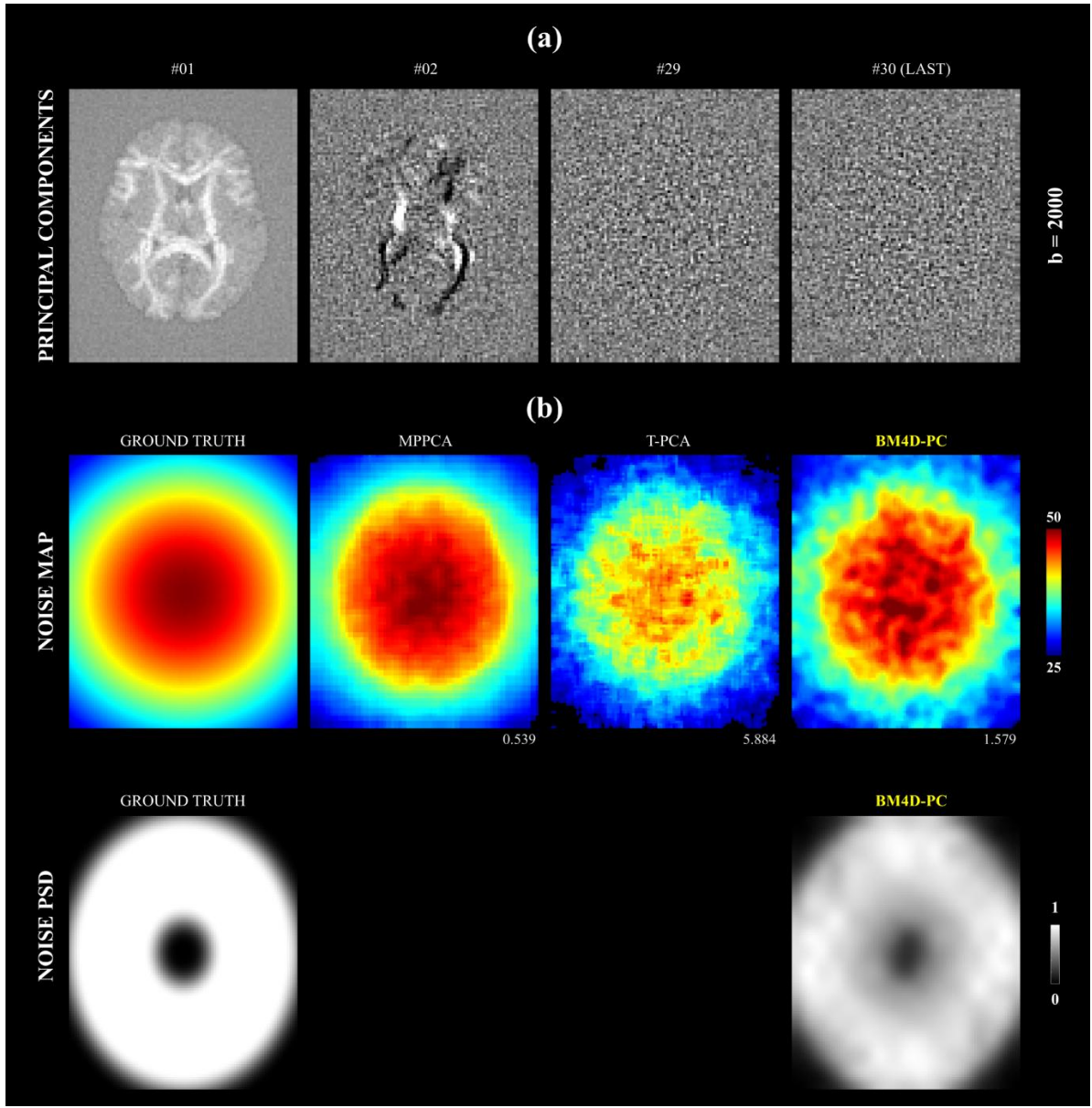

*Figure S1* – In silico experiment, colored noise at 5% noise level. (a) Example of global PCA for noise estimation, in which we show the Principal Components of the highest shell (b = 2000). Notice how the last two components are majorly comprised by noise. Therefore, they are used for estimating the noise map and PSD. (b) Top row – Noise maps. The numbers on the bottom right represent the RMSE of each method with respect to the ground truth; Bottom row – Noise PSD. They are normalized for visualization purposes. Note how the estimated PSD contains the overall shape of the ground truth.

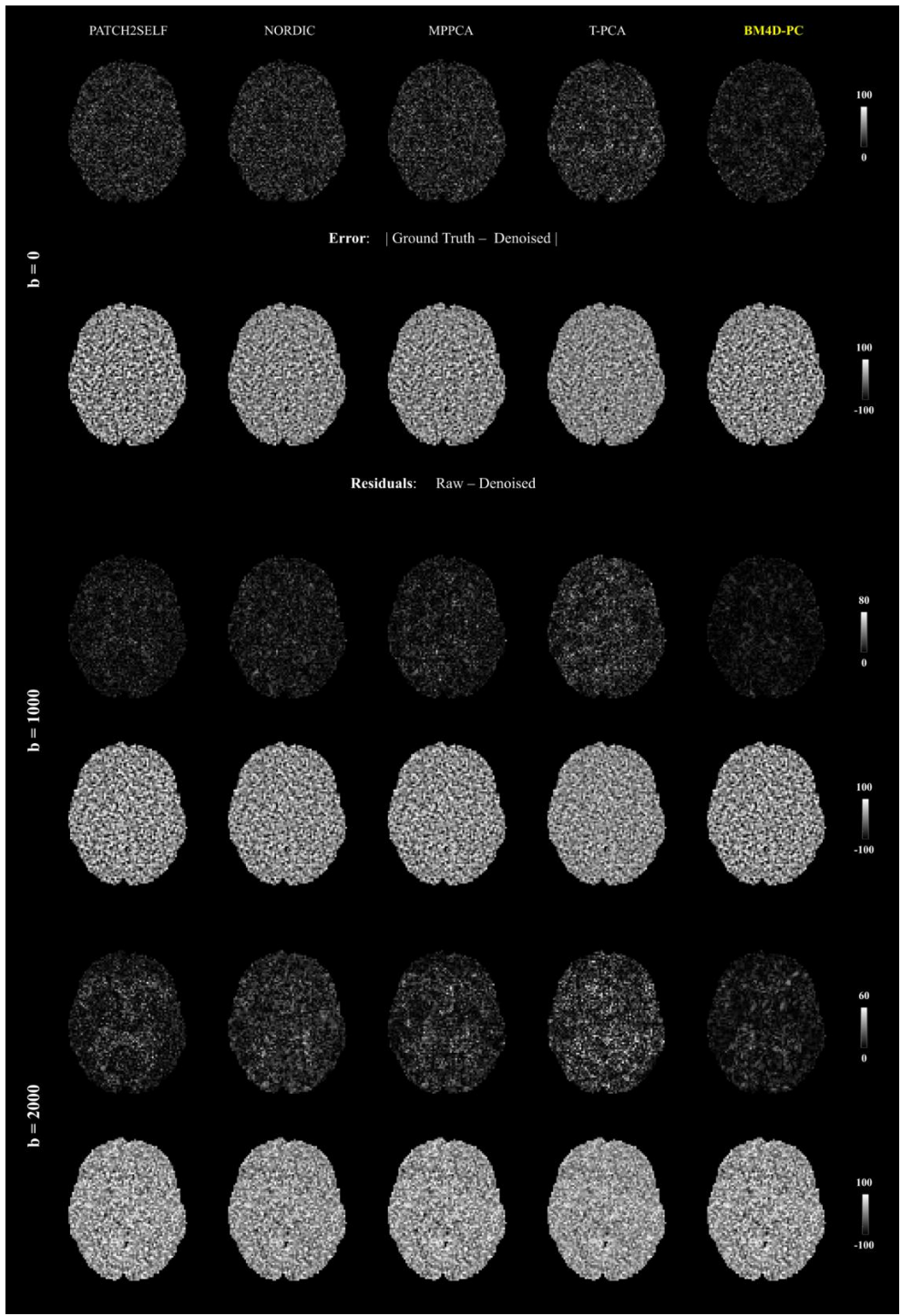

*Figure S2* – In silico experiment, colored noise at 5% noise level. For each b-value, we show: (top) the error, which is the absolute voxel-wise difference between the Ground Truth image and the corresponding denoising result; (bottom) the residuals, which are the difference between the Noisy image and the corresponding denoising result.

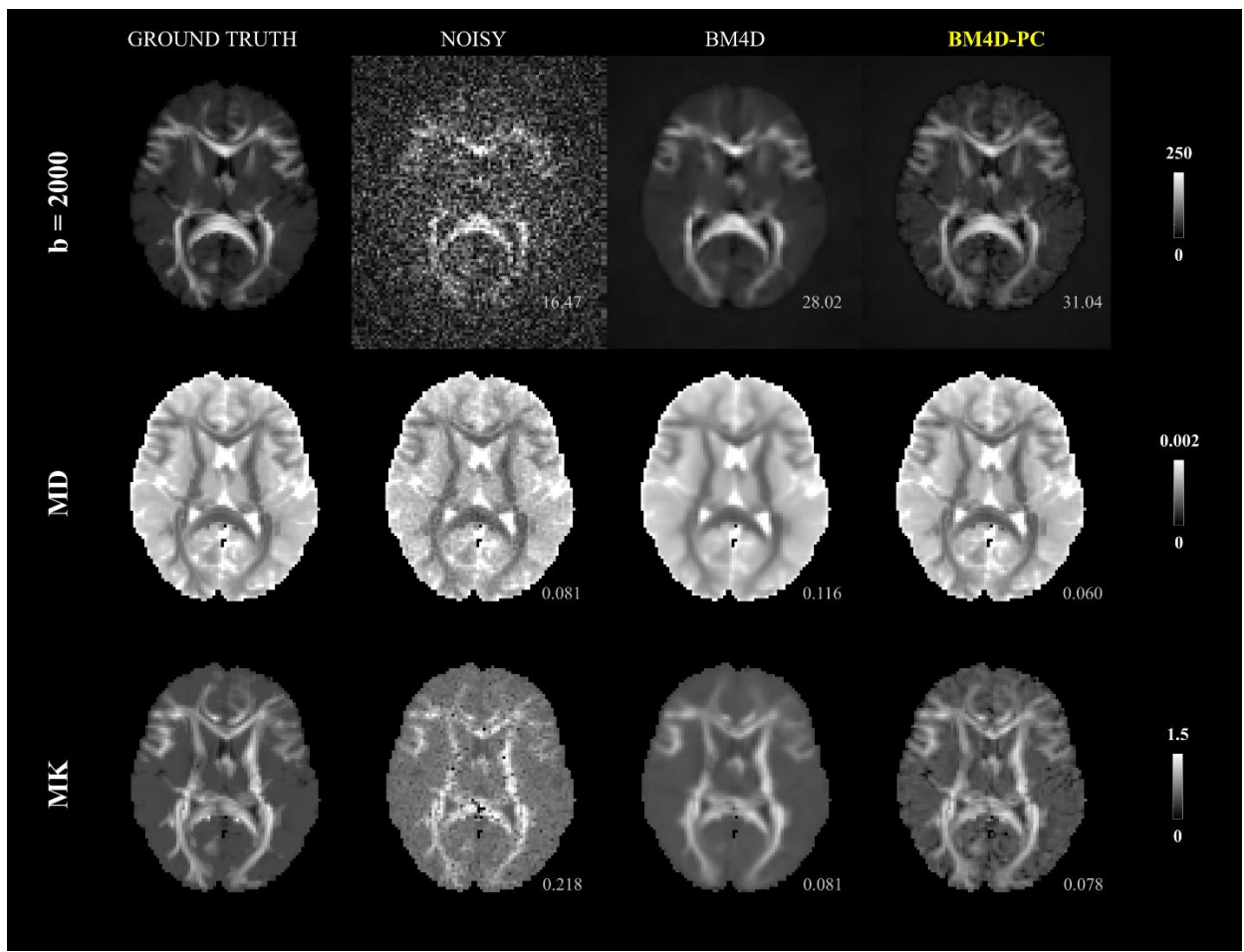

*Figure S3* - In silico experiment, colored noise at 5% noise level. BM4D-PC, compared with BM4D (applied to each DWI volume in the image domain (Top) Results of a representative slice of a single DWI (bvalue = 2000), DTI metric FA (middle), and DKI metric MK (bottom).The numbers on each image represent the PSNR (dB) (for the DWI) and RMSE (for the diffusion metrics).

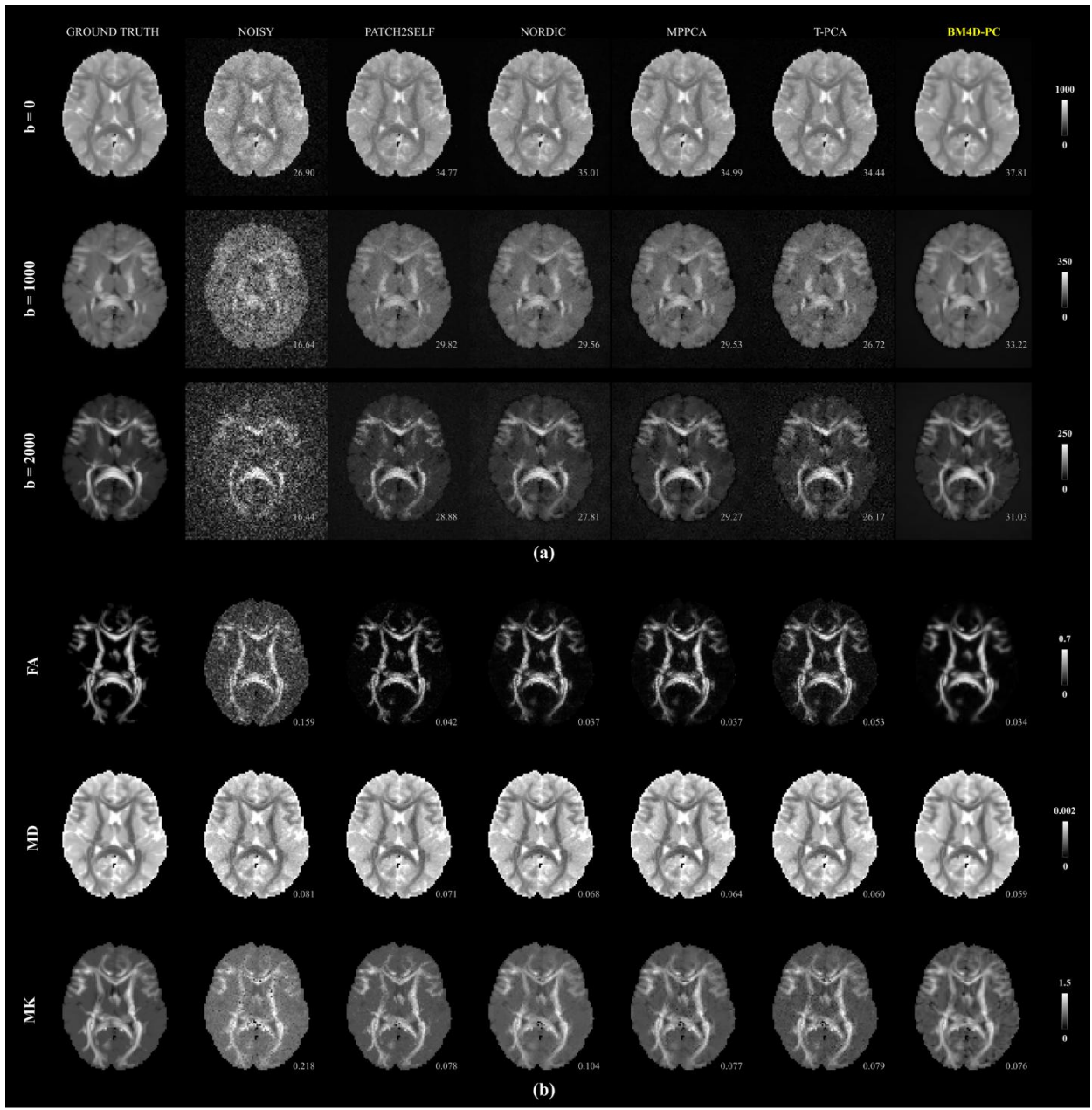

***Figure S4*** - *In silico experiment, white noise at 5% noise level. (a) Results of a representative slice of a single DWI, in which we show Ground Truth and Noisy images, followed by the denoised version of each method. The numbers on each image represent the PSNR (dB) of the 3D volume. (b) DTI metrics FA, MD, and DKI metric MK . The numbers on each image represent the RMSE with respect to the Ground Truth.*

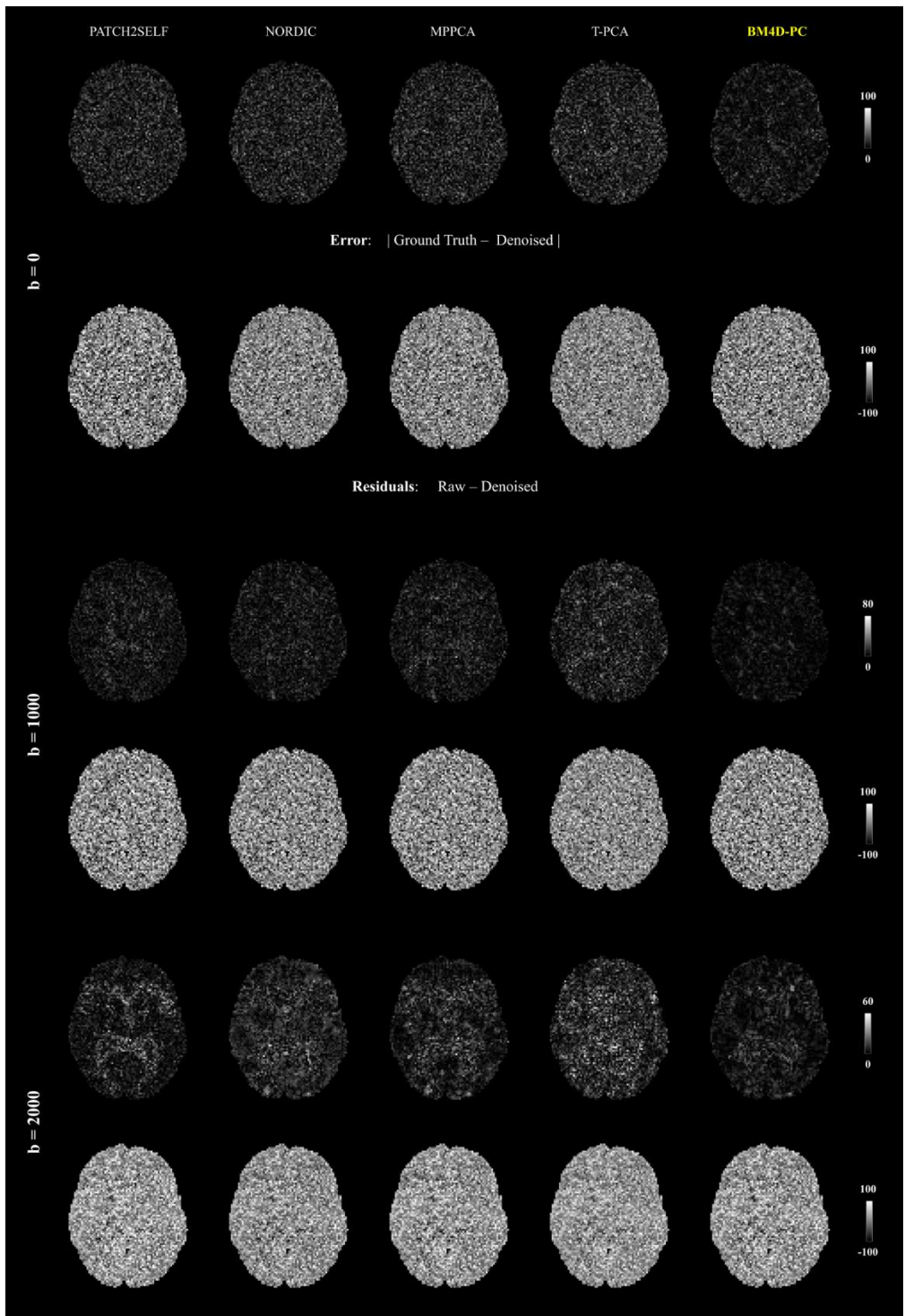

***Figure S5*** - *In silico experiment, white noise at 5% noise level. For each b-value, we show: (top) the error, which is the absolute voxel-wise difference between the Ground Truth image and the corresponding denoising result; (bottom) the residuals, which are the difference between the Noisy image and the corresponding denoising result.*

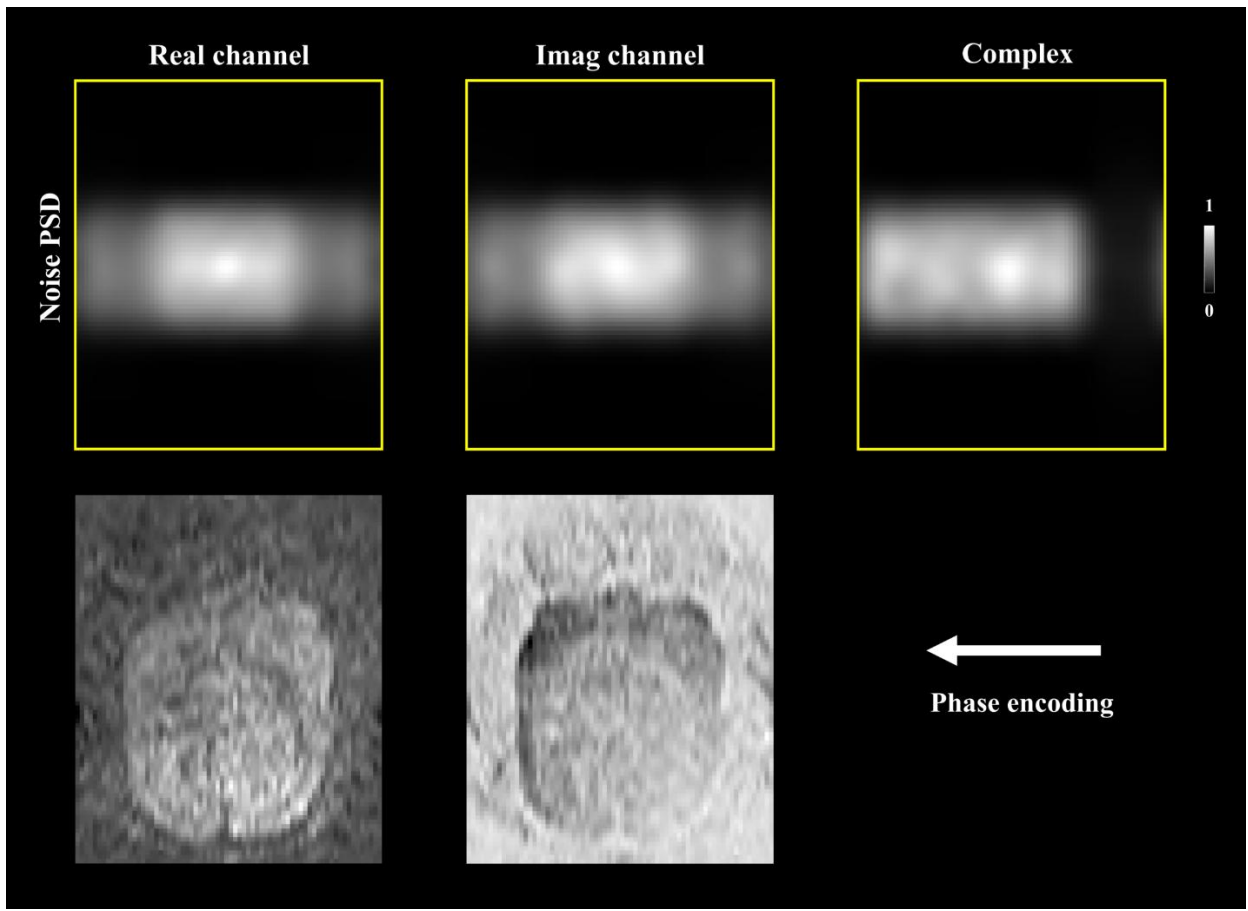

*Figure S6* – Noise PSD estimation test using the Marmoset data. (top) Noise PSDs obtained when using only Real channel (left), only Imaginary channel (middle) and Complex data (right). (bottom) Real and Imaginary images are shown for reference. For this PSD estimation test, we did not perform phase stabilization. Note that when using either the Real or the Imaginary channel, separately, the estimated noise PSD is conjugate-symmetric. When Complex data is used (both channels together), the expected non-symmetric partial Fourier sampling pattern becomes evident. Importantly, the partial Fourier was performed on the phase encoding dimension, whereas zero-filling was done on the frequency encoding dimension.

|  |  | NOISE LEVEL 1% | | | NOISE LEVEL 5% | | | NOISE LEVEL 10% | | |
|---|---|---|---|---|---|---|---|---|---|---|
|  |  | b = 0 | b = 1k | b = 2k | b = 0 | b = 1k | b = 2k | b = 0 | b = 1k | b = 2k |
| **PSNR** | NOISY | 40.86 | 31.03 | 30.79 | 26.87 | 17.22 | 17.00 | 20.89 | 11.53 | 10.16 |
|  | PATCH2SELF | 47.88 | 41.29 | 40.22 | 34.76 | 30.34 | 29.08 | 29.17 | 24.24 | 21.18 |
|  | NORDIC | 48.60 | 41.78 | **40.96** | 35.01 | 29.99 | 28.29 | 29.29 | 24.37 | 20.29 |
|  | MPPCA | 48.54 | 41.73 | 40.84 | 35.00 | 30.01 | 29.68 | 29.13 | 24.85 | 23.23 |
|  | T-PCA | 47.85 | 40.21 | 39.10 | 34.44 | 27.21 | 26.65 | 28.71 | 21.58 | 20.66 |
|  | BM4D-PC | **49.54** | **42.90** | 40.70 | **37.78** | **33.45** | **31.46** | **33.19** | **29.36** | **24.62** |
|  |  |  |  |  |  |  |  |  |  |  |
| **SSIM** | NOISY | 0.984 | 0.853 | 0.809 | 0.754 | 0.324 | 0.257 | 0.512 | 0.159 | 0.085 |
|  | PATCH2SELF | 0.997 | 0.984 | **0.983** | 0.940 | 0.857 | 0.834 | 0.830 | 0.635 | 0.465 |
|  | NORDIC | 0.997 | 0.986 | **0.983** | 0.944 | 0.842 | 0.841 | 0.830 | 0.652 | 0.604 |
|  | MPPCA | 0.997 | 0.986 | **0.983** | 0.945 | 0.846 | 0.854 | 0.829 | 0.658 | 0.653 |
|  | T-PCA | 0.997 | 0.980 | 0.969 | 0.940 | 0.752 | 0.692 | 0.824 | 0.515 | 0.429 |
|  | BM4D-PC | **0.998** | **0.989** | 0.982 | **0.975** | **0.932** | **0.905** | **0.938** | **0.880** | **0.798** |

***Table S1*** *– In silico experiment for white noise. Quantitative metrics of the DWIs are separated by noise level and b-value. Best results in bold.*

|  |  | NOISE LEVEL 1% | NOISE LEVEL 5% | NOISE LEVEL 10% |
|---|---|---|---|---|
| **FA** | NOISY | 0.031 | 0.159 | 0.266 |
|  | PATCH2SELF | 0.011 | 0.042 | 0.082 |
|  | NORDIC | **0.010** | 0.037 | 0.064 |
|  | MPPCA | **0.010** | 0.037 | 0.061 |
|  | T-PCA | 0.012 | 0.053 | 0.097 |
|  | BM4D-PC | 0.011 | **0.034** | **0.050** |
| **MD** $(\times 10^{-3})$ | NOISY | 0.015 | 0.081 | 0.199 |
|  | PATCH2SELF | 0.014 | 0.071 | 0.138 |
|  | NORDIC | 0.014 | 0.068 | 0.142 |
|  | MPPCA | 0.014 | 0.064 | 0.124 |
|  | T-PCA | **0.012** | 0.060 | 0.117 |
|  | BM4D-PC | 0.015 | **0.059** | **0.104** |
| **MK** | NOISY | 0.038 | 0.218 | 0.456 |
|  | PATCH2SELF | 0.024 | 0.078 | 0.176 |
|  | NORDIC | 0.022 | 0.104 | 0.236 |
|  | MPPCA | 0.022 | 0.077 | 0.167 |
|  | T-PCA | **0.021** | 0.079 | 0.167 |
|  | BM4D-PC | 0.030 | **0.076** | **0.160** |

***Table S2*** – *In silico experiment for white noise. Root Mean Squared Error (RMSE) values for the diffusion metrics. Best results in bold.*

| Dataset | Method | Main parameters |
|---|---|---|
| Insilico | PATCH2SELF | patch_radious=0;<br>model = MLPRegressor(activation='relu',<br>hidden_layer_sizes=(64,64,64,64),<br>learning_rate_init=3e-3,<br>early_stopping=False,<br>max_iter=500) |
| | NORDIC | ARG.kernel_size_PCA=[5,5,5] |
| | MPPCA | kernel = [5 5 5]<br>step = [3 3 3]<br>shrinkage = threshold<br>algorithm = jespersen |
| | T-PCA | patch_radius = [2,2,2]; step_size = 2 |
| | BM4D-PC | **Hard-Thersholding stage**<br>N1 = [4 4 4];% block size<br>N2 = 16; % max number of similar blocks<br>Ns = [5 5 5];% radious of search window<br><br>**Wiener stage**<br>N1_wiener = [4 4 4];% block size<br>N2_wiener = 32; % max num of similar blocks<br>Ns_wiener = [5 5 5];% radius of Search Window |
| EDDEN | PATCH2SELF | ** Same for all ** |
| | NORDIC | ARG.kernel_size_PCA=[7,7,7] |
| | MPPCA | kernel = [7 7 7]; step = [4 4 4] |
| | T-PCA | patch_radius = [3,3,3]; step_size = 2 |
| | BM4D-PC | ** Same for all ** |
| uFA | PATCH2SELF | ** Same for all ** |
| | NORDIC | ARG.kernel_size_PCA=[5,5,5] |
| | MPPCA | kernel = [5 5 5]; step = [3 3 3] |
| | T-PCA | patch_radius = [2,2,2]; step_size = 2 |
| | BM4D-PC | ** Same for all ** |
| Marmoset | PATCH2SELF | ** Same for all ** |
| | NORDIC | ARG.kernel_size_PCA=[7,7,7] |
| | MPPCA | kernel = [7 7 7]; step = [4 4 4] |
| | T-PCA | patch_radius = [3,3,3]; step_size = 2 |
| | BM4D-PC | ** Same for all ** |

*Table S3* – *Main parameters of the denoising methods*

**Comments with respect to the parameters:**

**BM4D-PC**: The parameters used in our implementation are based on the profile 'np' (normal profile) provided by default on BM4D. There are also two more profiles already setup and available to use, the 'lc' (low complexity) and 'mp' (modified profile), which will be less and more "aggressive" respectively.

**MPPCA, NORDIC, and T-PCA:** The patch sizes were selected based on the number of voxels within each patch, ensuring that the total number of voxels was the smallest possible value that still exceeded the number of diffusion-weighted volumes. Step size for MPPCA is automatically calculated by the Designer algorithm [1]. We added a step size for T-PCA and set it to 2 for the sake of speed.

**PATCH2SELF:** We tried several options and none of the linear regressors (OLS, Lasso, Ridge) would provide good results. Thus, we used a Multilayer Perceptron regressor. This was used by Kang et al. [2], although with different parameters. Please check https://github.com/B9Kang/Multidimensional-Self2Self-MD-S2S.git .

1. Chen J, Ades-Aron B, Lee HH, et al. Optimization and validation of the DESIGNER preprocessing pipeline for clinical diffusion MRI in white matter aging. *Imaging Neurosci*. 2024;2:1-17.

2. Kang B, Lee W, Seo H, Heo HY, Park H. Self-supervised learning for denoising of multidimensional MRI data. *Magn Reson Med*. 2024;92(5):1980-1994.